\documentclass{article}

\usepackage{amssymb}

\def\a{{\alpha}}
\def\b{{\beta}}

\def\l{{\lambda}}
\def\m{{\mu}}

\def\on#1#2{\mathop{\vbox{\ialign{##\crcr\noalign{\kern2pt}
$\scriptstyle{#2}$\crcr\noalign{\kern2pt\nointerlineskip}
\kern-2pt$\hfil\displaystyle{#1}\hfil$\crcr}}}\limits}

\def\nn{ \nonumber }
\def\bq{ \begin{equation} }
\def\eq{ \end{equation} }
\def\ben{ \begin{eqnarray} }
\def\en{ \end{eqnarray} }
\def\frac#1#2{{#1\over #2}}
\def\dfrac#1#2{{\displaystyle{#1\over#2}}}

\newtheorem{prop}{Proposition}
\begin{document}
\title{The Maupertuis principle
  and integrable systems.}
\author{
A.V. Tsiganov\\
\it\small Department of Mathematical and Computational Physics,
 Institute of Physics,\\
\it\small  St.Petersburg University,
 198 904, St.Petersburg, Russia.\\
\it\small e-mail: tsiganov@mph.phys.spbu.ru}
 \date{}
\maketitle


\begin{abstract}
We discuss some special classes of canonical transformations of the
extended phase space, which relate integrable systems with a common
Lagrangian submanifold. Various parametric forms of trajectories are
associated with different integrals of motion, Lax equations,
separated variables and action-angles variables. In this review we
will discuss namely these induced transformations instead of the
various parametric form of the geometric objects.
\end{abstract}
\vskip2cm

\section{Introduction}
\setcounter{equation}{0} Let us begin with a simple example.
Consider an ellipse defined by the standard implicit equation
\[\dfrac{x^2}{a^2}+\dfrac{y^2}{b^2}=1\,.\]
One can represent this ellipse by the parametric equations
\bq
x=a\,\sin(t)\,,\qquad y=b\,\cos(t)\,,\qquad t\in[0,2\pi]\,.
\label{ellipse1}
\eq
It is known, there are infinitely many parameterisations of a given
curve. For instance, we can reparameterise an ellipse by using another
parameter
\bq
\widetilde{t}=(a^2+b^2)\,t+(a^2-b^2)\,\sin(t)\,.
\label{ellipse2}
\eq
Construction of different polynomial, rational and other
parameterisations of the plane curves is a subject of classical
algebraic geometry.

In classical mechanics the same ellipse may be identified with
integral trajectories of various integrable systems on the common
phase space. In this case the parameter $t$ is the time variable
conjugated to some Hamilton function $H$. As an example, the first
parametric form of the ellipse (\ref{ellipse1}) is related to the
two-dimensional oscillator, while the second parameterisation
(\ref{ellipse2}) may be associated to the Kepler model.

In this and several other unexpected situations in mathematics,
dynamics is occasionally invading mathematical objects in which time is
not present in the definition and yet the object can be endowed with
various dynamical system structures, continuous or discrete.

Thus, there is a problem of finding suitable parameterisation of
curves, which are associated with various integrable systems. What do
we know about this problem?

We consider a mechanical system defined by some Lagrangian function
${\cal L}(q,\dot{q})$ or a Hamilton function $H(p,q)$ on the
$2n$-dimensional phase space $\cal M$ with local coordinates
$\{q_j,p_j\}_{j=1}^{n}$.  According to Maupertuis' principle the
extremals $q_j=\gamma_j(t)$ of the action functional
\bq
{\cal S}=\int_A^B {\cal
L}\,\Bigl(\,q(t),\dot{q}(t)\Bigr)\,dt\,,\qquad q=(q_1\,\ldots,q_n)
\label{acts}
\eq
coincide with the extremals of the reduced action functional
${\cal S}_0$ on the fixed energy surface
\bq
{\cal Q}^{2n-1}=\Bigl(\,H(p,q)=E\,\Bigr)\,.
\label{fen}
\eq
Recall that the Maupertuis' principle was at first enunciated in 1744
\cite{mau50,lag88,jac36}.  The modern interpretation of the
Maupertuis' principle may be found in \cite{arn89,nov82,bkf94}. The
reduced action ${\cal S}_0=\int p\,dq$ is independent of any
evolution parameter. Moreover, even its initial parameter $t$ and the
corresponding Hamilton function cannot be restored from the reduced
action problem \cite{arn89}. Nevertheless, solutions of the
corresponding variational problem are the initial trajectories
$q_j=\gamma_j(t)$ in the common nonparametric form \cite{arn89}. For
instance, trajectories of the Kepler system are conic sections
$a\,r^{-1}=1+b\,\cos\phi$, which may be ellipses, hyperbolas or
parabolas at $E<0$, $E>0$ and $E=0$, respectively.

So, any integral trajectory may be parameterised by using another
parameter, $\widetilde{t}$,such that
\[q_j=\gamma_j(t)=\beta_j(\widetilde{t})\,,\qquad t\in
[A,B]\,,\quad \widetilde{t}\in [C,D]\,.
\]
According to Maupertuis' principle all the initial trajectories have
one nonparametric form on the surface ${\cal Q}^{2n-1}$. Therefore,
for all these trajectories we could introduce common parametric form
$q_j=\beta_j(\widetilde{t})$ and a new local Hamilton function
$\widetilde{H}$ defined on ${\cal Q}^{2n-1}$. Two main problems are
how to find the new parametric form of the initial trajectories and
how to obtain the new Hamilton function defined on the whole phase
space ${\cal M}$.

Consider the corresponding Hamilton-Jacobi equation
\[
\dfrac{\partial {\cal S}}{\partial t}+
H\left(\dfrac{\partial{\cal S}}{\partial q_1},
\ldots,\dfrac{\partial{\cal S}}{\partial q_n}\right)=0\,.
\]
In the invariant geometric Hamilton-Jacobi theory \cite{vk77} any
hyperplane ${\cal Q}^{2n-1}$ in $\cal M$ may be called the
Hamilton-Jacobi equation. Its solution is an $n$-dimensional
Lagrangian submanifold ${\cal C}^{(n)}$ in ${\cal M}$, such that
${\cal C}^{(n)}\subset{\cal Q}^{2n-1}$. By definition, a Lagrangian
submanifold is one where the symplectic form $\Omega$ vanishes when
restricted to it, i.e. $\left.\Omega\right|_{\cal C}=0$. This
definition is completely invariant with respect to change of local
coordinates and parametric representations of the Lagrangian
submanifold \cite{arn89,vk77}.  As above, we could consider various
parameterisations of a given Lagrangian submanifold. Each new
parametric form yields a new Hamiltonian system related to the same
geometric surface.

Below we consider integrable systems on $\cal M$ with $n$ integrals
in involution. According to the Liouville theorem \cite{arn89} for
any integrable system the corresponding $n$-dimensional Lagrangian
submanifold depends at least of $n$-arbitrary constants.  So
integrability is a geometric property and  it does not depend on the
choice of the parameterisation of the Lagrangian surface. Starting
with a known Lagrangian surface of some integrable system we can try
to get new integrable models by using various parametric forms of
this surface. In this case, we can expect that the initial and
resulting integrable systems have a lot of common properties. The
main problem is how to find different parameterisations and the
corresponding sets of integrals of motion defined on the whole phase
space $\cal M$. Generally there is no rule for how to proceed. Each
case is different.

Usually the Lagrangian submanifold depends on $m\geq n$ arbitrary
constants. The $n$ constants $\alpha_1,\ldots,\alpha_n$ are
identified with the values of integrals of motion $I_j=\alpha_j$
\cite{vk77}, while the remaining $m-n$ constants
$a_1,\ldots,a_{m-n}$ are free parameters.  In this case we have a
freedom related to the choice of $n$ integrals of motion from the
$n+m$ initial constants of motion. This freedom permits us to
associate a family of integrable systems with one Lagrangian surface.
In this review we discuss a special class of different parametric
forms of a given surface, which is associated with mutual
permutations of energy $E$ and arbitrary parameter $a_k$. The
corresponding reparameterisation of ${\cal C}^{(n)}$ we will call
the generalized Kepler change of time.

The aim of this paper is to bring together some old and new examples
of integrable systems related with the various parametric forms of
the one Lagrangian surface. The passage from a given
parameterisation to the another one gives rise to the
transformations of all the properties of integrable systems, such as
integrals of motion, Lax equations, separated variables and the
action-angles variables. In this paper we discuss namely these
induced transformations instead of the various parametric form of
the geometric objects.

The initial symplectic form $\Omega$ is equal to zero on the
Lagrangian submanifold ${\cal C}^{(n)}$. We can suppose, that the
space $\cal M$ may be equipped with another form
$\widetilde{\Omega}$, which is equal to zero on the same surface
${\cal C}^{(n)}$. In this case for a given Lagrangian surface time
$t$ and symplectic structure may be changed simultaneously.
Moreover, we can try to embed this $n$-dimensional surface into the
various phase spaces. For instance, the $n$-dimensional Kepler
problem may be identified with the geodesic flow on the sphere $S^n$
\cite{lc06,mos70}. The Kolosoff transformation maps the Kowalevski
top into the two-dimensional St\"ackel system \cite{kol01}. The same
top may be related with the geodesic motion on $SO(4)$ \cite{avm89}
or with the Neumann system on the sphere $S^2$ \cite{hh87}. Here we
will not discuss such composed transformations of time and phase
space.

By embedding a Lagrangian submanifold ${\cal C}^{(n)}$ into the
infinite-dimensional phase space, we can identify  ${\cal
C}^{(n)}$ with an invariant manifold of some hierarchy of
nonlinear evolution equations. Such finite dimensional manifolds
are invariant with respect to the action of all flows of the
hierarchy and they are naturally expected to be integrable since
all the flows of hierarchy commute on them. For instance, given
Lagrangian submanifolds may be realised via stationary or
restricted flows. Here we will not discuss relation of
finite-dimensional integrable systems with soliton equations and
restrict ourselves to finding new parameterisations of the known
trajectories only.

Below we will consider many well known mechanical systems, such as
the St\"{a}ckel systems, Toda lattices, H\'enon-Heiles and Holt
systems, integrable systems with the quartic potential  and the
Goryachev-Chaplygin top. The results we present are, for the most
part, not new and we do not provide detailed proofs (these can be
found in the papers cited). What may be new and interesting is an
exposition of canonical transformations of the extended phase space
as different parametric forms of integrable geometric objects and
the action of these transformations on the properties of integrable
systems.

\section{The Maupertuis-Jacobi transformations}
\setcounter{equation}{0}
Let $M$ be an $n$-dimensional Riemannian manifold with the metric
$g_{ij}$. On the cotangent bundle ${\cal M}=T^*M$ consider a
Hamiltonian system with the natural Hamilton function $H(p,q)$
\bq
H(p,q)=T(p,q)+V(q)=\sum_{i,j}^n g_{ij}(q)\,p_i\,p_j+V(q)\,.
\label{nham}
\eq
On the smooth submanifold ${\cal Q}^{2n-1}$ (\ref{fen}), integral
trajectories of the Hamiltonian vector field
$\xi={\textsf{sgrad}}\,H(p,q)$ coincide with integral trajectories
of another vector field
$\widetilde{\xi}={\textsf{sgrad}}\,\widetilde{H}(p,q)$, where the
new Hamilton function is given by
\bq
\widetilde{H}(p,q)=\widetilde{T}(p,q)=
\sum_{i,j}^n \dfrac{g_{ij}(q)}{E-V(q)}\,p_i\,p_j\,.
\label{mopham}
\eq
Integral trajectories have two different parametric forms
$q_j=\gamma_j(t)=\beta_j(\widetilde{t})$ on the surface ${\cal
Q}^{2n-1}$ only. However, the resulting Hamilton function
describes geodesic motion and, therefore, on the whole phase
space we can determine the so-called Maupertuis transformation
\[\xi \mapsto\widetilde{\xi},\] which relates the initial hamiltonian
vector field $\xi$ on $\cal M$ with the other hamiltonian vector
field $\widetilde{\xi}$ defined on the same phase space $\cal M$
\cite{bkf94}.

If $\widetilde{t}$ be the time along trajectories of the new vector
field $\widetilde{\xi}$, then the Maupertuis mapping  gives rise to
so-called Jacobi transformations \cite{jac36,lanc49,rosp95} of the
Hamilton function (\ref{mopham}) and of the time variable
\bq
d\widetilde{t}=\Bigl(\,E-V(q)\,\Bigr)\,dt\,.
\label{moptime}
\eq
This Jacobi transformation explicitly describes new parameterisation of
the common integral trajectories $q(t)=q(\widetilde{t})$ and determines
the new Hamilton function $\widetilde{H}(p,q)$ (\ref{mopham}).

The Maupertuis transformation maps any integrable system with a
natural Hamilton function $H(p,q)$ into the other integrable system
on the same phase space $\cal M$. Namely this property has been used
for the search of the new integrable systems (see references within
\cite{lag88,jac36,bkf94,rosp95,sel98}). The Maupertuis principle for
integrable systems with a nonnatural Hamilton function is discussed
in \cite{nov82}.

The property of integrability is independent on the choice of
parametric form of trajectories. However, some criteria of
integrability drastically depend upon the parameterisation. For
instance, the method of singularity analysis associates
integrability with the Kowalevski-Painlev\'{e} property, i.e. the
only singularities of the solutions of the equations of motion are
movable poles $(t-t_0)^{-m}$ in the complex $t$-plane
\cite{avm89,rgb89}. There exist cases of integrable systems with the
rational integrals of motion, whose analytic structure permits
solutions with algebraic singularities of the type $(t-t_0)^{-m/k}$,
($k$ being a positive integer larger than one). These systems
satisfy to the so-called "weak" Painlev\'{e} property (see review
\cite{rgb89}).

We have some examples of integrable systems related with one
Lagrangian submanifold  \cite{rgb89,ts98b,ts98c}, which satisfy to
the usual Kowalevski-Painlev\'e property and the "weak" Painlev\'{e}
property, respectively. These systems are related with the different
parametrisations of one geometric object. So we can see that a change
of parametric form leads to a change of the Kowalevski-Painlev\'e
criteria of integrability.

\section{Canonical transformation of the extended \\ phase space}
\setcounter{equation}{0}

To find new parameterisation of the known integral trajectories or
of the Lagrangian surfaces one has to introduce a new parameter
$\widetilde{t}$ and the corresponding Hamilton function
$\widetilde{H}$. Thus, to describe explicitly the following mapping
\bq
t\mapsto \widetilde{t}\,,\qquad H(p,q)\mapsto \widetilde{H}(p,q)\,
\label{timetr}
\eq
we extend initial phase space $\cal M$ by adding to it the new
coordinate $q_{n+1}=t$ with the corresponding momentum $p_{n+1}=-H$.
The resulting $2n+2$-dimensional space ${\cal M}_E$
\cite{lanc49,syng60} is the so-called extended phase space of the
hamiltonian system. We emphasize that $H(p,q)$ is the Hamilton
function on $\cal M$, but $H$ is an independent variable in the space
${\cal M}_E$. The energy $E$ is a fixed value of the variable $H$ or
the function $H(p,q)$.

To describe evolution on the extended phase space ${\cal M}_E$ we
introduce the generalized Hamilton function \cite{lanc49,syng60}
\bq
{\cal
H}(p_1,\ldots,p_{n+1};q_1\,\ldots,q_{n+1})=H(p,q)-H\,.\label{oham}
\eq
The Hamilton equations for the variables $q_{n+1}=t$ and $p_{n+1}=-H$
are
\[\dfrac{dt}{d\tau}=1\,,\qquad\dfrac{dH}{d\tau}=0\,.\]
Here $\tau$ is a generalized time (parameter) associated to the
generalized Hamilton function $\cal H$. The time variable $t$ is a
cyclic coordinate and the conjugated momentum is a constant of
motion. Other $2n$ equations coincide with the initial Hamilton
equations on the zero-valued energy surface
\bq
{\cal H}(p,q)=H(p,q)-H=0\,.\label{constr}
\eq
Thus, our initial hamiltonian system on $\cal M$  may be immersed
into the hamiltonian system on ${\cal M}_E$. Using this immersion and
canonical transformations of the extended phase space ${\cal M}_E$
\cite{lanc49,syng60} we introduce transformations
\[\begin{array}{ccccc}
  \quad  &(H,t)&
  \mbox{\rm\small canonical transformations}
    & (\widetilde{H},\widetilde{t})&\quad \\
    {\cal M}_E & \quad & \longrightarrow & \quad & {\cal M}_E \\
    \uparrow & \quad & \quad & \quad & \downarrow \\
    {\cal M} & \quad & \quad & \quad & {\cal M} \\
    \uparrow & \quad & \quad & \quad & \downarrow \\
    H(p,q)   & \quad & \quad & \quad & \widetilde{H}(p,q)
  \end{array}\]
which map an initial Hamilton function $H(p,q)\mapsto
\widetilde{H}(p,q) $ into another Hamilton function defined on the
same phase space. Of course the similar mapping $t\mapsto
\widetilde{t}$ permits us to describe new parameterisation of the
corresponding integral trajectories.

Note two different classical definitions of the canonical
transformations are known \cite{arn89}.
\par\noindent
1.{\it  Canonical transformations preserve the canonical form of the
Hamilton-Jacobi equations}.
\par\noindent
2.{\it Canonical transformations preserve the differential 2-form,
$\Omega=\sum_{i=1}^n dp_i\wedge dq_i$, on $\cal M$}.
\par\noindent
For instance the first definition is used in textbooks on
variational principles of classical mechanics
\cite{jac36,lanc49,syng60,shar66}. The second definition was later
introduced to consider geometry of the phase space
\cite{arn89,arng85,vk77}.

Below we use the first definition of canonical transformations
because the Maupertuis-Jacobi transformation and the Kepler change
of the time preserve the canonical form of the Hamilton-Jacobi
equation, but it retains the corresponding differential 2-form
$\Omega=\sum_{i=1}^{n+1} dp_i\wedge dq_i$ on the level ${\cal
Q}^{2n-1}$ (\ref{fen})  only \cite{bkf94}.

We introduce general canonical transformations of the extended phase
space ${\cal M}_E$
\ben &t\mapsto\widetilde{t}\,,\qquad &d\widetilde{t}=v(p,q)\,dt\,,\nn\\
\label{ttr}\\
&H\mapsto\widetilde{H}\,,\qquad &\widetilde{H}=v(p,q)^{-1}\,H\,.\nn
\en
which change the initial equations of motion
\[
\dfrac{dq_i}{d\widetilde{t}}=
v^{-1}(p,q)\left(\,\dfrac{dq_i}{dt}-\widetilde{H}\,\dfrac{\partial\,v}{\partial
p_i}\,\right)\,,\qquad
\dfrac{dp_i}{d\widetilde{t}}=
v^{-1}(p,q)\left(\,\dfrac{dp_i}{dt}+\widetilde{H}\,\dfrac{\partial\,v}{\partial
q_i}\right)\,,
\]
but preserve the canonical form of the Hamilton-Jacobi equation
\bq
\dfrac{\partial {\cal S}}{\partial t}+H=0\,,\qquad\mbox{\rm
where}\qquad {\cal S}=\int \Bigl(\,p\,dq-H\,dt\,\Bigr)\,.
\label{hjeq}
\eq
and retain the corresponding zero-energy surfaces (\ref{constr}) at
$v(p,q)\neq 0$
\[\widetilde{\cal H}(p,q)=v(p,q)^{-1}{\cal H}(p,q)=0\,.
\]
Zeroes of the function $v(p,q)$ determine the behavior of the system
with respect to the inversion of time \cite{shar66}. Here we do not
consider this problem in detail.

The Maupertuis-Jacobi mapping (\ref{mopham}), (\ref{moptime}) may be
rewritten as such a canonical transformation (\ref{ttr}) of the
extended phase space
\[T(p,q)\mapsto\widetilde{T}(p,q)=v(p,q)^{-1}\,T(p,q)\,,\qquad
v(p,q)=E-V(q)\,,
\]
which maps an initial geodesic flow into another geodesic flow. This
map preserves integrability, if the function $v(p,q)$ is constructed
by any potential $V(q)$, which may be added to the initial kinetic
energy $T(p,q)$ without loss of integrability \cite{jac36}.

In contrast with the Maupertuis-Jacobi transformations, even if the
general canonical transformation of ${\cal M}_E$ (\ref{ttr}) retains
integrability, we have no general method to construct new integrals
of motion starting with initial ones. To solve this problem in
\cite{ts98b,ts98c,ts99c,ts99d,ts00a}, we used some analogies with
the one known example of such transformations due to Kepler.

\section{The Kepler change of the time}
\setcounter{equation}{0}

We begin with brief description of the Kepler change of the time
\cite{lc06}. We commence with two-dimensional oscillator defined by
the Hamilton function
\[H_{osc}(p,q)=p_1^2+p_2^2+a\,(q_1^2+q_2^2)+b\,,\qquad a,b\in{\mathbb R}.\]
For this system the Kepler canonical transformation (\ref{ttr}) of
${\cal M}_E$ with the function
\bq
v(p,q)=q_1^2+q_2^2 \label{vkepl}
\eq
preserves integrability. After change of the time (\ref{ttr}) and
the point canonical transformation to other variables
\bq
x=q_1q_2\,,\qquad y=(q_1^2-q_2^2)/2\,\label{keplvar}
\eq
integral trajectories of the oscillator come to be trajectories of
the Kepler problem defined by
\bq
\widetilde{H}_{kepl}(p,x)=\dfrac{H_{osc}(p,q)}{q_1^2+q_2^2}
=p_x^2+p_y^2+\dfrac{b}{2\sqrt{x^2+y^2}}+a\,.\label{keptr}
\eq
Various parametric and nonparametric forms of the common
trajectories are discussed in \cite{lc06,arn89,arng85}. Coincidence
of the integral trajectories may be regarded as a local result,
whereas the corresponding canonical transformation of the extended
phase space preserves integrability in the whole initial phase space.

As for the Maupertuis-Jacobi transformation, the function $v(p,q)$
(\ref{vkepl}) in the Kepler transformation (\ref{keptr}) could be
identified with the oscillator potential $V(q)$. Below we prove that
it is a simple coincidence. Nevertheless canonical transformations
of the type (\ref{ttr}) have been called the coupling constant
metamorphoses in \cite{hgdr84,rgb89} because of this casual
coincidence.

The Kepler change of time has been generalized by Liouville
\cite{li49}. The Liouville integrable systems are systems with the
natural Hamilton function
\[\widetilde{H}(p,q)=\widetilde{T}+\widetilde{V}\,,\]
 where the kinetic and potential energies are given by
\[
\widetilde{T}=v^{-1}(q)\,\sum_{i=1}^n a_i\,p_i^2\,,\qquad
\widetilde{V}=v^{-1}(q)\,\sum_{i=1}^n U_i\,,\qquad
v(q)=\sum_{i=1}^n v_i\,.
\]
The functions $a_i$, $v_i$ and $U_i$ depends only on the variable
$q_i$.

For the Liouville system the following quantities
\[\widetilde{I}_j=a_j\,p_j^2+U_j-\widetilde{H}\,v_j\,,\qquad j=1,\ldots,n\,,\]
are integrals of motion in involution and $\sum I_j=0$. Thus, we
have $n$ quadratic integrals of motion including the Hamilton
function and consequently the Liouville systems are completely
integrable. Note the two-dimensional oscillator and the Kepler model
belong to the Liouville family of integrable systems.

We recall how equations of motion were integrated in quadratures by
Liouville \cite{li49}. From ${\widetilde I}_j=\a_j$, one obtains a
system of differential equations
\[
\dfrac{dq_j}{\sqrt{a_j\,(\a_j+\widetilde{E}\,v_j-U_j)}}
=\dfrac{d\widetilde{t}}{v(q)}\,,\qquad j=1,\ldots,n\,.
\]
Here $\widetilde{H}=\widetilde{E}$ and time $\widetilde{t}$ is
associated with the Hamilton function $\widetilde{H}$. Choosing a
new time variable ${t}$ according to
\[dt=v^{-1}(q)\,\widetilde{t}\]
we come down to the system of equations
\[\dfrac{dq_j}{\sqrt{a_j\,(\a_j+\widetilde{E}\,v_j-U_j)}}
=d{t}\,.
\]
It allows one to find integral trajectories $q_j=\gamma_j(t)$. After
that the new parameter $t$ may be expressed in terms of the initial
time $\widetilde{t}$ by the quadrature
\bq
\widetilde{t}=\int^t v(\,q_1(\tau),\ldots,q_n(\tau)\,)\,d\tau\,.
\label{lioutime}
\eq
This transformation is related to a new parametric form of the same
trajectories $q_j=\beta_j(\widetilde{t})$.

In fact, Liouville tacitly used canonical transformation of the
extended phase space (\ref{ttr}) and considered a new integrable
Hamilton function $H=v(q)\,\widetilde{H}$ instead of the initial one.
After integration of equations of motion for the new system we have
change parametric form of the trajectories in order to describe
solution of the initial problem. For instance in parametric form of
the ellipse the Kepler time (\ref{ellipse2}) is explicitly the
Liouville quadrature (\ref{lioutime}).

Note canonical transformations of the extended phase space
(\ref{ttr}) have a natural counterpart in quantum mechanics. Namely,
it is known that the standard eigenvalue problem of the Hamiltonian
operator $H(p,q)$,
\[
H(p,q)\,\Psi=(H_0+a\,V+b)\,\Psi=E\,\Psi\,,
\]
may be associated with the eigenvalue problem of the charge
operator $a$
\[
\widetilde{H}(p,q)\,\widetilde{\Psi}=
(\widetilde{H}_0+(b-E)\,
V^{-1})\,\widetilde{\Psi}=-a\,\widetilde{\Psi}\,.
\]
In quantum mechanics such a duality of the two eigenvalue problems
has been used by Schr\"odinger  and many other \cite{rids82}.
Canonical transformation of the extended phase space (\ref{ttr}) is
an analogue of this duality. It is interesting that for the first
time this duality has been studied for the quantum Kepler problem. In
the Birman-Schwinger formalism function $v(q)$ is called a "sandwich"
potential \cite{rids82}. Recall that in the Birman-Schwinger
formalism we can estimate the spectrum and eigenfunctions of the one
Hamiltonian, $\widetilde{H}$, by using the known spectrum and
eigenfunctions of the dual Hamiltonian, $H$. Moreover, for some
quantum models it is a single known way to find solutions of the
initial Schr\"odinger equation. Below we briefly discuss a similar
property for the quantum Toda lattice.

\section{The generalized Kepler change of time}
\setcounter{equation}{0}

The Maupertuis-Jacobi mapping is traditionally used for the search of
new integrable systems. The Kepler change of time and the Liouville
reparameterisation of trajectories have been used for integration of
equations of motion. In this Section we propose some generalisations
of the Kepler-Liouville results, which may be useful for the search
of new integrable systems as well.

All the Liouville systems are particular case of St\"ackel
integrable systems. Therefore let us briefly recall some necessary
facts about  St\"{a}ckel systems \cite{st95}. The nondegenerate
$n\times n$ St\"{a}ckel matrix ${\bf S}$, its $j$-th column of which,
$s_{kj}$, depends only on $q_j$
\[\det {\bf S}\neq 0\,,\qquad \dfrac{\partial s_{kj}}{\partial q_m}=0\,,
\quad j\neq m
\]
defines the set of functionally independent integrals of motion,
$\{I_k\}_{k=1}^n$, where
\bq
I_k=\sum_{j=1}^n c_{jk}\left(p_j^2+U_j\,(q_j)\,\right)\,,
\qquad c_{jk}=\dfrac{{\bf S}^{kj}}{\det{\bf S}}\,,
\label{int1}
\eq
which are quadratic in the momenta. Here ${\bf C}=[c_{jk}]$ denotes
the inverse matrix of ${\bf S}$ and  ${\bf S}^{kj}$ is the cofactor
of the element $s_{kj}$.

\begin{prop} \cite{ts98b}
If the two St\"{a}ckel matrices ${\bf S}$ and $\tilde{{\bf S}}$ be
distinguished by the $m$-th row only, i.e.
\[s_{kj}=\tilde{s}_{kj}\,,\qquad k\neq m\,,\]
the corresponding Hamilton functions $I_m$ and $\widetilde{I}_m$
(\ref{int1})  with a common set of potentials $U_j$  are related by
canonical transformations of the extended phase space ${\cal M}_E$
\bq
 {I}_m \mapsto \widetilde{I}_m=
v^{-1}(q)\,I_m(p,q)\,,\qquad d\widetilde{t}_m=v(q)\,dt_m
\,, \label{ntrans1}
\eq
where
\bq
v(q_1,\ldots,q_n)=
\dfrac{\det\widetilde{{\bf S}}(q_1,\ldots,q_n)}{\det{{\bf S}}(q_1,\ldots,q_n)}\,.
\label{dham}
\eq
\end{prop}
The proposed generalization of the Kepler transformation maps one
integrable St\"ackel system into another integrable St\"ackel system.
For instance, the St\"ackel  matrices for the oscillator and the
Kepler problem are
\bq
{\bf S}=\left({\begin{array}{cc} 1 & 1 \\ 1 &
-1\end{array}}\right)\,,\qquad
\widetilde{{\bf S}}=\left({\begin{array}{cc} q_1^2 & q_2^2
\\ 1 & -1\end{array}}\right)
\,.\label{sh}
\eq
It is obvious that the Kepler change of time (\ref{keptr}) coincides
with the proposed mapping (\ref{ntrans1}).

We return to the Kepler transformation (\ref{keptr}) of ${\cal
M}_E$. After permutation of coordinates and momenta
$(q_{1,2}\leftrightarrow p_{1,2})$ the Hamilton function for the
Kepler problem
\[\widetilde{H}_{kepl}=a\,\dfrac{p_1^2+p_2^2+a^{-1}
(q_1^2+q_2^2)+a^{-1}b}{p_1^2+p_2^2}
=a\,\dfrac{H_{osc}}{H_{free}}
\]
becomes a ratio of the Hamilton function, $H_{osc}$, for the
oscillator and the Hamilton function, $H_{free}$, for the free
motion.

So for any two integrable systems the ratio of their Hamilton
functions could be the Hamilton function of a third integrable system
on the same phase space. The main remaining problem is a search of a
complete set of integrals of motion.

We consider two integrable hamiltonian systems on the common phase
space ${\cal M}$. These systems are defined by the two sets of
independent integrals of motion, $\{I_j\}_{j=1}^n$, and
$\{J_j\}_{j=1}^n$, in involution, i.e.
\[ \{I_j,I_k\}=0\,\qquad {\rm and} \qquad
\{J_j,J_k\}=0\,,\quad j,k=1,\ldots,n\,.
\]
Introduce the antisymmetric matrix ${\cal K} =({\bf I}\otimes {\bf
J})$, which is the inner product of the two independent vectors of
integrals ${\bf I}$ and ${\bf J}$ in ${\mathbb R}^n$. Any column or
row of this matrix defines a set of $n-1$ independent functions
\[{\cal K}_{ij}=({\bf I}\otimes {\bf J})_{ij}
=I_i\,J_j-I_j\,J_i\,,\qquad i,j=1,\ldots,n\,.
\]

\begin{prop}\cite{ts99c}
If all the differences of integrals of motion $(I_j-J_j)$ with the
common index $j=1,\ldots,n$ are in involution, i.e.
\bq
\Bigl\{I_j-J_j\,,I_k-J_k\Bigr\}=0\,,\qquad j,k=1,\ldots,n\,,\label{usl}
\eq
then the ratio of integrals
\bq
K_m=\dfrac{I_m}{J_m}
\label{newh}
\eq
and $n-1$ functions $K_j$, $j\neq m$
\bq K_j= \dfrac{{\cal K}_{mj}}{J_m}=\dfrac{I_m\,J_j-I_j\,J_m}{J_m}=
 K_m\,J_j-I_j\,,\qquad
m\neq j=1,\ldots,n
\label{newi}
\eq
are integrals of motion for new integrable system on the same phase
space.
\end{prop}
Thus the mapping (\ref{newh}) defines a canonical transformation
(\ref{ttr}) of the extended phase space, which preserves the
property of integrability. To apply this transformation we have to
find two integrable systems satisfying condition (\ref{usl}).

We consider a pair of the St\"{a}ckel systems with a common
St\"{a}ckel matrix ${\bf S}$ and with different potentials $U_j$.
Namely, in addition to the system with integrals $\{I_k\}$
(\ref{int1}), we introduce the second integrable system with the
similar integrals of motion,
\bq
J_k=\sum_{j=1}^n c_{jk}\Bigl(\,p_j^2+W_j\,(q_j)\,\Bigr)\,,\quad
k=1,\ldots,n\,.
\label{int2}
\eq
At least one potential $U_j\,(q_j)$ has to be functionally
independent of the corresponding potential $W_j\,(q_j)$.

\begin{prop}\cite{ts99c}
Any two integrable systems defined by the same St\"{a}ckel matrix
${\bf S}$ and by functionally independent potentials $U_j\,(q_j)$
and $W_j\,(q_j)$ satisfy the necessary condition (\ref{usl}) of
the previous proposition. Thus, the ratio of the two St\"{a}ckel
integrable Hamiltonians defines new integrable system
\bq
(I_m,J_m)\quad\mapsto\qquad
K_m=v(p,q)^{-1}\,I_m=\dfrac{I_m}{J_m}\,.\label{ntrans2}
\eq
\end{prop}
It is obvious that all the integrals $I_k$ and $J_k$ differe by the
potential part
\[(I_k-J_k)=\sum_{j=1}^2 c_{jk}\left[\,U_j(q_j)-W_j(q_j)\,\right]\]
depending on the coordinates  $q$ only.  Thus systems with a common
St\"{a}ckel matrix ${\bf S}$ satisfy condition (\ref{usl}).

The Hamilton function (\ref{ntrans2}) has the following form
\bq
H(p,q)=K_m=\dfrac{ \sum_{j=1}^n c_{jm}\left[p_j^2+U_j\,(q_j)\right]
-\b_m}{\sum_{j=1}^n c_{jm}\left[p_j^2+W_j\,(q_j)\right] }\,,
\qquad\b_m\in{\mathbb R}\,.
\label{newhs}
\eq
This Hamiltonian $H(p,q)$ is a rational function in the momenta, but
next one can try to use canonical transformations to simplify it.
Occasionally, one obtains again a natural type of Hamilton function.
For instance, according to \cite{ts99c}, integrable systems with the
following Hamilton functions
\ben
H_{I}&=&p_x^k\,p_y^k+ a\,\bigl(x\,y\bigr)^{-
\textstyle{{k}\over{k+1}} }\,,\qquad
a\,,k\in{\mathbb R}\,,\nn\\
\label{gfk}\\
H_{II}&=&p_x^k+p_y^k+ a\,\bigl(x\,y\bigr)^{-
\textstyle{{k}\over{k+1}} }\,,\nn
\en
belong to the proposed family of generalized St\"ackel systems. At
$k=1$ the first Hamiltonian coincides with the Hamiltonian of the
Kepler problem. At $k=2$ the second integrable Hamiltonian has been
found by Fokas and Lagerstr\"{o}m  (see references within
\cite{rgb89}). In this case both initial systems satisfy to
Kowalevski-Painlev\'{e} criterion, whereas the resulting
Fokas-Lagerstrom system admits asymptotic solutions with fractional
powers in $t$ \cite{rgb89}.

\section{Properties of the change of time for the St\"ackel systems}
\setcounter{equation}{0}
For the St\"ackel family of integrable systems we proposed two
different examples, (\ref{ntrans1}) and (\ref{ntrans2}), of the
canonical transformations (\ref{ttr}) of the extended phase space
${\cal M}_E$. Now we consider some properties of these
transformations.

Recall that the common level surface of the St\"ackel integrals $I_j$
\bq
M_\a=\left\{z\in {\mathbb R}^{2n}: I_i(z)=\a_i\,,~i=1,\ldots,n\right\}
\label{subm}
\eq
is diffeomorphic to the $n$-dimensional real torus. We can
immediately construct the one-dimensional separated equations
\bq
p_j^2=\left(\dfrac{\partial {\cal S}_0}{\partial
q_j}\right)^2=P_j(q_j)= \sum_{i=1}^n \a_i
s_{ij}(q_j)-U_j(q_j,a)\,,\qquad a_k\in{\mathbb R}\,. \label{curvj}
\eq
Here ${\cal S}_0$ is a reduced action functional \cite{st95}.
Integral trajectories $q_j(t,\a_1,\ldots,\a_n)$ are determined from
the following equations
\bq
\sum_{j=1}^n\int\dfrac{s_{kj}(q_j)\, dq_j}
{\sqrt{P_j(q_j)}}=\b_k\,,\qquad\b_1=t\,.\label{stinv}
\eq
In fact the polynomial $P(\l)$ and the contour of integration depend
upon the values, $\a_j$, of the integrals of motion, which are
dropped in the notation.

For the rational entries of $\bf S$ and rational potentials $U_j(q_j)$
the Riemann surfaces (\ref{curvj}) are isomorphic to the hyperelliptic
curves
\bq
{\cal C}_j:\quad \mu_j^2=P(\l_j)=\sum_{k=1}^{2g+1}
a_k\,\lambda_j^k\,. \label{sthc}
\eq
Considered together these curves determine the $n$-dimensional
Lagrangian submanifold in the phase space ${{\cal M}=\mathbb R}^{2n}$
\bq
{\cal C}^{(n)}:\qquad {\cal C}_1(p_1,q_1)\times{\cal C}_2(p_2,q_2)
\times\cdots\times{\cal C}_n(p_n,q_n)\,, \label{lagr}
\eq
which is decomposed into plane curves. Applying Arnold's method
\cite{arn89} we find that the action variables have the form
\bq
{\mathfrak s}_j=\oint_{A_j} \sqrt{P(\l_j)\,}\,d\l_j\,.\qquad
\label{stact}\\
\eq
The Abel transformation linearizes the equations of motion on the
Lagrangian submanifold ${\cal C}^{(n)}$ in terms of abelian
differentials of the first kind on the corresponding spectral curves
\cite{ts97d,ts98b}.

We consider a pair of the St\"ackel systems related by the first
generalization of the Kepler mapping (\ref{ntrans1}) such that the
potentials $U_j(\l)=\sum a_k\l^k$ depend on  arbitrary parameters
$a_k$. According to \cite{ts98b} initial and resulting integrable
systems are associated with algebraic hyperelliptic curves
(\ref{sthc}) ${\cal C}$ and $\widetilde{\cal C}$ described by
\ben
&{\cal C}:\qquad &\m^2=\sum a_j\,\l^{j}+a_m\l^{m}+ E\l^k +
\sum \a_i\l^i\,,\nn\\
\label{sturm}\\
&\widetilde{\cal C}:\qquad &\m^2=\sum a_j\,\l^j+\widetilde{E}\l^{m}+
a_k\l^k + \sum
\widetilde{\a}_i\l^i\,.\nn
\en
Here the $n$ coefficients $\{\a_j\}$ and $\{\widetilde{\a}_j\}$ are
values of the integrals of motion such that $E=\a_1$ and
$\widetilde{E}=\widetilde{\a}_1$. The other coefficients
$a_j\in{\mathbb R}$ are arbitrary parameters (charges), which define
the potential part of the Hamilton function $H(p,q)=T(p)+V(q,a)$.
Canonical transformations of ${\cal M}_E$ give rise to mutual
permutation of the energy $E$ and one of the parameters $a_k$
(charge).

The initial and resulting Riemann surfaces are topologically
equivalent. One can prove that the corresponding integrable systems
are topologically equivalent too. Moreover, initial curves coincide
with the resulting curves at the special values of integrals of
motion
\[E=a_m\,,\qquad\widetilde{E}=a_k\,,
\qquad \a_j=\widetilde{\a}_j\,,\quad j=2,\ldots,n\,.
\]
In this case we have two different parametric forms of the common
Lagrangian submanifold ${\cal C}^{(n)}$ (\ref{lagr}) depending on
$n$ values of integrals $\a_j$ and constants $a_k$. On the
corresponding submanifolds  ${\cal M}_\a$ and $\widetilde{\cal
M}_\a$ (\ref{subm}) integral trajectories of the initial system,
$q_j(t)$ (\ref{stinv}), coincide with integral trajectories of the
resulting system, $q_j(\widetilde{t})$. In the neighbourhood of the
intersection of these submanifolds we can introduce the common set
of the action-angle variables (\ref{stact}). In this region the
Kepler transformation (\ref{ntrans1}) retains the differential 2-form
$\Omega$ in ${\cal M}_E$ and the function $v(p,q)=v({\mathfrak s})$
depends on the common action variables only.

We consider a pair of the St\"ackel systems related by the second
generalization of the Kepler mapping (\ref{ntrans2}). According to
\cite{ts99c} initial and resulting integrable systems are associated
with algebraic hyperelliptic curves (\ref{sthc}) ${\cal C}$ and
$\widetilde{\cal C}$ of the form
\[
{\cal C}:\qquad \m^2=\sum a_i\,\l^i+ \sum {\a}_j\l^j\quad
\mapsto\quad
\widetilde{\cal C}:\qquad \m^2=\sum b_k\,\l^k+ \sum
\widetilde{\a}_m\l^m\,.\nn
\]
Here the resulting coefficients are rational functions of the initial
ones \cite{ts99c}. So canonical transformations of ${\cal M}_E$
(\ref{ntrans2}) give rise to transformations of the modulus of the
curves only.

As mentioned above the corresponding Riemann surfaces and integrable
systems are topologically equivalent. At the special choice of
integrals $\{\a,\,\widetilde{\a}\}$ and parameters of potentials
$\{a_i,\,b\}$ integral trajectories of the initial system coincide
with trajectories of the resulting system. In a small region of $\cal
M$ this generalization of the Kepler transformation (\ref{ntrans2})
preserves the differential 2-form $\Omega$ in ${\cal M}_E$ and the
function $v(p,q)$ depends on the action variables only.

In an attempt to understand the origin of the conservation of
integrability by canonical transformation of ${\cal M}_E$
(\ref{ttr}) one can attempt to rewrite equations of motion in the
Lax form
\[\{H(p,q),L(\l)\}=\Bigl[\,L(\l),A(\l)\,\Bigr]\,.\]
For the some classes of St\"ackel systems the Lax matrices have been
constructed in \cite{ts97d,ts98b,ts98d}.

We consider some examples only. One of the simplest Lax matrix
$L(\l)$ for the one-dimensional St\"ackel systems is given by
\bq
L(\lambda)=\left(\begin{array}{cc}
  p & \lambda-q \\
  \\
  \left[\dfrac{\phi(\l)}{\lambda-q}\right]_{MN} & -p
\end{array}\right)\,.
\label{block}
\eq
Here $\phi(\lambda)=\sum \phi_k\,\l^k$ is a parametric function of
the spectral parameter $\lambda$ and the elements, $[z]_{MN}$, are
the linear combinations of the Taylor projections
\bq
{[ z ]_{N}}=\left[\sum_{k=-\infty}^{+\infty} z_k\l^k\,
\right]_{N}\equiv
\sum_{k=0}^{N} z_k\l^k\,,
\label{cutmn}
\eq
or the Laurent projections \cite{ts96b}. The coefficients $\phi_k$ of
the function $\phi(\lambda)$ are the constants of motion, which may
be parameters $a_k$ or integrals of motion. So at arbitrary $M,N$
the family of the Lax matrices $L_\phi(\l)$ may be associated with a
web of algebraic curves instead of one concrete curve.

For instance, we present some one-dimensional Hamilton functions, the
functions $\phi(\lambda)$ and the corresponding Lax matrices
associated with the first Taylor projection $[z]_{01}$. An
application of the pure numeric function $\phi(\lambda)$ yields
\ben
&&H=p^2+a\,q^2+b\,q\,,\qquad
\phi=-a\,\lambda^2-b\,\lambda\,,\nn\\
\nn\\
&&L=\left(\begin{array}{cc} p& \lambda-q\\ -a\,(\lambda+q)-b & -p
\end{array}\right)\,,\qquad
A=\left(\begin{array}{cc} 0&1\\ -a & 0\end{array}\right)\,.
\label{laxosc}
\en
The corresponding spectral curve is given by
\[{\cal C}:\qquad \mu^2=a\,\lambda^2+b\,\lambda-H\,.\]
Note that in the spectral curve we can substitute integrals of
motion, while in the corresponding Lagrangian submanifold we have to
substitute the values of integrals of motion.

From (\ref{sturm}) the first possible change of this curve looks like
\[{\cal C}\quad\mapsto\quad
\widetilde{\cal
C}:~\mu^2=a\,\lambda^2+(b-\widetilde{H})\,\lambda+c\,.
 \]
The associated canonical transformation of the extended phase space,
\[
\widetilde{H}=v^{-1}\,(H+c)=q^{-1}\,(H+c)\,,\qquad
\]
changes the Lax matrices by the following rule
\bq
\label{ltr}
\phi=-a\,\lambda^2-(b-\widetilde{H})\,\lambda\,,\qquad
\widetilde{L}=L+\widetilde{H}\,\left(\begin{array}{cc}
  0 &0 \\  1&0\end{array}\right)
  \,,\qquad \widetilde{A}=v^{-1}(q)\,{A}\,.
\eq
The second possible change of the curve
\[{\cal C}\quad\mapsto\quad
\widehat{\cal C}:~\mu^2=(a-\widehat{H})\,\lambda^2+b\,\lambda+c\]
may be related to the other canonical transformation of the extended
phase space,
\[
\widehat{H}=v^{-1}\,(H+c)=q^{-2}\,(H+c)\,,\]
such that
\ben
&&\phi=-(a-\widehat{H})\,\lambda^2-b\,\lambda\,,\nn\\
\label{ltr1}\\
&&\widehat{L}=L+\widehat{H}\,\left(\begin{array}{cc}
  0 &0 \\  \l+q&0\end{array}\right)
  \,,\qquad \widehat{A}=v^{-1}(q)\,\left[\,
  {A}+\widehat{H}\,\left(\begin{array}{cc}
  0 &0 \\  1&0\end{array}\right)\,\right]\,.
\nn
\en
It is not hard to check that the Poisson bracket relations for all
these Lax matrices $L(\l)$, $\widetilde{L}(\l)$ and $\widehat{L}(\l)$
are closed into the linear $r$-matrix algebra. At the first case the
$r$-matrix is a constant matrix $r=\Pi/(\l-\mu)$, whereas the second
and third $r$-matrices depend on dynamical variables. Here $\Pi$ is
a permutation of auxiliary spaces \cite{ft87}.

We turn now to the original change of time in Kepler problem
(\ref{keptr}). The Lax matrices for the two-dimensional oscillator,
\bq
{\cal L}(\lambda)=
\left(\begin{array}{cc}
  L_1(\lambda,p_1,q_1) & 0 \\ \\
  0 & L_2(\lambda,p_2,q_2)
\end{array}\right)\,,\qquad
{\cal A}(\lambda)= \left(\begin{array}{cc}
  A_1(\lambda) & 0 \\ \\
  0 & A_2(\lambda)
\end{array}\right)\,,
\label{drlax}
\eq
may be constructed from two independent $2\times 2$ blocks
$L_j(\lambda)$ and $A_j(\lambda)$ of (\ref{laxosc}). The
corresponding spectral curve ${\cal C}={\cal C}_1\times{\cal C}_2$
is a product of two hyperelliptic curves.

The Kepler mapping (\ref{keptr}) gives rise to the following
transformation of the Lax matrices
\ben
\widetilde{\cal L}(\lambda)&=&{\cal L}(\lambda)-
\widetilde{H}_{kepl}\left(\begin{array}{cccc}
  0 & 0 & 0 & 0 \\
  \l+q_1 & 0 & 0 & 0 \\
  0 & 0 & 0 & 0 \\
  0 & 0 & \l+q_2 & 0
\end{array}\right)\,,\nn\\
\label{4tr}\\
\widetilde{\cal A}(\l)&=&v(p,q)^{-1}\,\left[{\cal
A(\l)}-\widetilde{H}_{kepl} \left(\begin{array}{cccc}
  0 & 0 & 0 & 0 \\
  1 & 0 & 0 & 0 \\
  0 & 0 & 0 & 0 \\
  0 & 0 & 1 & 0
\end{array}\right)\right]\,.
\nn
\en
The spectral curve $\widetilde{\cal C}$ remains a product of two new
hyperelliptic curves. The initial oscillator and the resulting
Kepler model are separable in the common variables, which lie on
these curves. Note that these separated variables are cartesian
coordinates for the oscillator and parabolic coordinates for the
Kepler problem.

Similar transformations of the Lax matrices (\ref{ltr}),
(\ref{ltr1}), (\ref{4tr}) may be proposed for the other
two-dimensional St\"ackel systems separable in cartesian or
parabolic coordinates.  For the two-dimensional St\"ackel systems
separable in  elliptic or polar coordinates transformation of the
Lax matrices has a more complicated form \cite{ts98b}. These
transformation may be constructed by using two different outer
automorphisms of infinite-dimensional representations of underlying
$sl(2)$ algebra proposed in \cite{ts98d}.

In the next Sections we consider similar transformations of the
Lax matrices (\ref{ltr}) and the spectral curves (\ref{sturm})
for integrable systems associated with non-hyperelliptic
algebraic curves and for the non-St\"ackel integrable systems.

\section{On integrable systems with quartic potential}
\setcounter{equation}{0}
According to \cite{rgb89,ts98c,ts00a}, canonical transformations of
the extended phase space relate three integrable cases of
Hen\'on-Heiles systems with the three integrable cases of the Holt
systems. One of these systems admits a $2\times 2$ Lax matrix and the
corresponding spectral curve is an hyperelliptic algebraic curve.
Another two systems possess $3\times 3$ Lax matrices and the
corresponding spectral curves are the trigonal algebraic curves
$\mu^3=P(\l)$. As for the St\"ackel systems transformations of these
$3\times 3$ Lax matrices are shifts of their entries by the element
of the extended phase space (\ref{ltr}).

We consider two integrable systems with quartic potential for which
$4\times 4$ Lax matrices were constructed in \cite{bef95} by applying
relations with stationary flows of some known integrable PDEs. The
first system belongs to the St\"ackel family and is separable in
cartesian coordinates. Its Hamilton function is
\bq H(p,q)=p_1^2+p_2^2+ \dfrac14\,\left(\,q_1^4
+6\,q_1^2\,q_2^2+q_2^4\,\right)
\eq
As above we could construct some block $4\times 4$ Lax matrices
(\ref{drlax}) for this system  in cartesian separated variables. The
corresponding spectral curve is a product of hyperelliptic curves.
The two pairs of the separated variables lie on these two curves.

Another $4\times 4$ Lax matrix for the same system may be obtained
from the Lax representation for the Hirota-Satsuma coupled KdV system
\cite{bef95}. This Lax matrix
\[
L(\l)=\left(\begin{array}{cccc} -p_1\,q_1&q_1^2& 0 & 1\\
\\
\l-{p_1^2}&{p_1\,q_1}&-\dfrac{q_1^2+q_2^2}{2}&0\\
\\
0 & \l &-{p_2\,q_2}& {q_2^2}\\
\\
-\l\dfrac{q_1^2+q_2^2}2&0&\l-p_2^2&p_2q_2
\end{array}\right)
\]
does not have a pure block-diagonal structure and the corresponding
spectral curve $\mu^4=P(\l)$ is not a product of two hyperelliptic
curves. The second Lax matrix $A(\l)$ may be found in \cite{bef95}.

For this St\"ackel system canonical transformation of the extended
phase space (\ref{ntrans1})
\[\widetilde{H}=(q_1^2+q_2^2)^{-1}\,(H-b)\]
gives rise to the shift of the first Lax matrix and the rescaling of
the second Lax matrix
\[\widetilde{L}=L+\widetilde{H}\,\left(\begin{array}{cccc}
  0 & 0 & 0 & 0 \\
  0 & 0 & 1 & 0 \\
  0 & 0 & 0 & 0 \\
  \l & 0 & 0 & 0
\end{array}\right)\,,
\qquad
\widetilde{A}=v^{-1}\,A\,.
\]
In contrast with the St\"ackel systems (\ref{sturm}),
transformation of the corresponding spectral curves acts on the
both side of the equation defining the curve
\bq
\begin{array}{lcccrc}
  {\cal C}:\qquad & \mu^4 & ~ & =\l^3 & -H\,\l^2 & +J\,\l\,, \\
  \\
  \widetilde{\cal C}:\qquad & \mu^4 & -2\mu^2\l\,\widetilde{H}
   & =\l^3 & -(\widetilde{H}^2-b)\,\l^2 & +\widetilde{J}\,\l\,.
\end{array}
\label{curvqu1}
\eq
Note that after canonical transformations of the other variables
$(p,q)$ the new Hamilton function $\widetilde{H}$ may be rewritten
in the natural form
\[\widetilde{H}=
p_x^2+p_y^2+\dfrac{x^2+2y^2-b}{2\,\sqrt{x^2+y^2}}\,.
\]

The second integrable system with the quartic potential is the
non-St\"ackel system defined by the following Hamilton function
\bq
H=p_1^2+p_2^2-\dfrac18\,\left(q_1^4+6\,q_1^2\,q_2^2+8\,q_2^4\,\right)\,.
\label{hqu2}
\eq
This system is separable after a so-called quasi-point canonical
transformation \cite{ts96a}.

The second system possesses $4\times 4$ Lax matrix
\[L=\left(\begin{array}{cccc}
\dfrac{q_2q_1^2}2+q_1p_1& -q_1^2& 2q_2& 2\\
\\
\dfrac{q_1^2q_2^2}{4}+p_1^2+q_1q_2p_1+\dfrac{\l}2&
-\dfrac{q_2q_1^2}2-q_1p_1& p_2+q_2^2+\dfrac{q_1^2}4& 0\\
\\ -2q_2\l& 2\l&-\dfrac{q_2q_1^2}2+q_1p_1& -q_1^2\\
\\
\l\left(-p_2+q_2^2+\dfrac{q_1^2}4\right)& 0&
\dfrac{q_1^2q_2^2}{4}+p_1^2-q_1q_2p_1+\dfrac{\l}2&
\dfrac{q_2q_1^2}2-q_1p_1
\end{array}\right)
\]
which was obtained using a gauge transformation of the Hirota-Satsuma
coupled KdV system \cite{bef95}.

For this system we propose the canonical transformation of the
extended phase space given by
\[\widetilde{H}=(q_1^2+4\,q_2^2)^{-1}\,(H-b)\,,\]
which preserves integrability. As for the St\"ackel systems,
transformation of the Lax matrices retains a shift of the first Lax
matrix and rescaling of the second Lax matrix
\[\widetilde{L}=L+2\widetilde{H}\,\left(\begin{array}{cccc}
  0 & 0 & 0 & 0 \\
  0 & 0 & 1 & 0 \\
  0 & 0 & 0 & 0 \\
  \l & 0 & 0 & 0
\end{array}\right)\,,\qquad
\widetilde{A}=v^{-1}\,A\,.
\]
The spectral curves are changed by the rule
\bq
\begin{array}{lcccrc}
  {\cal C}:\qquad & \mu^4 & ~ & =\l^3 & +4H\,\l^2 & +J\,\l\,, \\
  \\
  \widetilde{\cal C}:\qquad & \mu^4 & -8\mu^2\l\,\widetilde{H}
   & =\l^3 & +4(b-4\widetilde{H}^2)\,\l^2 & +\widetilde{J}\,\l\,.
\end{array}
\label{curvqu2}
\eq
Recall that for  St\"ackel systems with quadratic integrals of
motion canonical transformations of the extended phase space
(\ref{ntrans1}),(\ref{ntrans2}) preserve separated variables lying
on the common hyperelliptic curve ({\ref{sturm}).

For the H\'enon-Heiles systems and systems with quartic potentials
transformations of the trigonal spectral curve $\mu^3=P(\l)$
\cite{ts98c} and the curve $\mu^4=P(\l)$ (\ref{curvqu1},
\ref{curvqu2}) have a more complicated form. In this case separated
variables for the initial and resulting systems are different
\cite{ts98c}. It means that the proposed canonical transformation of
the extended phase space preserve integrability, but seriously
changes other properties of the systems.

\section{The Toda lattices.}
\setcounter{equation}{0}
Before proceeding further, it is useful to recall some known facts
about the Toda lattices (all details may be founded in the review
\cite{rs87}).

Let $\mathfrak g$ be a real, split, simple Lie algebra of ${\rm
rank}\,{\mathfrak g}=n$ and let $K(\cdot,\cdot)$ be its Killing form
and $P$ be a system of simple roots. we identify the phase space with
the coadjoint algebra ${\cal M}\simeq{\mathfrak g}_R^*={\mathfrak
g}_+^*\oplus{\mathfrak g}_-^*$. Here ${\mathfrak g}_+$ is a Borel
subalgebra  and ${\mathfrak g}_-$ is the opposite nilpotent
subalgebra of $\mathfrak g$.

The Lax matrices for the Toda lattices are
\ben
{\cal L}&=& \sum_{i=1}^n p_ih_i+\sum_{\a\in P} \cdot\exp K(\a,q)\cdot
e_{-\a}+\sum_{\a\in P} a_\a e_\a\,,\nn\\
\label{todal}\\
{\cal A}&=&-\sum_{\a\in P} \exp K(\a,q)\cdot e_{-\a}\,,\qquad
a_\a\in{\mathbb R}
\,.\nn
\en
The Hamilton function is given by
\bq
H(p,q)=\dfrac12K({\cal L},{\cal L})=\dfrac12K(p,p)+ \sum_{\a\in
P}\,a_\a\,e^{\a(q)}\,.
\label{todah}
\eq
Recall that in a shifted version of the Adler-Kostant-Symes scheme in
order to construct the Toda orbit (\ref{todal}) in ${\cal M}$ we have
to translate a dynamical orbit living in ${\mathfrak g}_+^*$ by
adding to it a constant vector $e=\sum a_\a e_\a$ from the remaining
part of $\cal M$. This vector  $e$ has to be a character and has to
be a constant.

We replace now the phase space $\cal M$ by the extended phase space
${\cal M}_E$. Roughly speaking, to consider the St\"ackel systems we
exchange the pure numerical function $\phi$ (\ref{laxosc}) by the
function with coefficients from ${\cal M}_E$ (\ref{ltr}),
(\ref{ltr1}). By a similar reasoning we try to construct the same
modification of the Adler-Kostant-Symes scheme. Namely we translate
the Toda orbit in ${\cal M}\simeq{\mathfrak g}_R^*$ by adding to it
a constant vector from the remaining part of the whole space ${\cal
M}_E$. As above this vector has to be a character and has to be a
constant with respect to the new time.

In addition we impose a constraint on the possible change of
parametric form of trajectories. As for the St\"ackel systems
(\ref{sturm}) the initial invariant polynomial has to generate the
arbitrary constant
\bq
K(\widetilde{\cal L},\widetilde{\cal L})=-b\,\label{newtod}
\eq
instead of the Hamilton function (\ref{todah}). This condition
together with the form of transformations of the Lax matrices
dictates to us a very special choice of the functions $v(p,q)$ in
(\ref{ttr}) for the Toda lattices.

\begin{prop}\cite{ts99c}
For each simple root $\b\in P$ and for any constant $b_\b\in{\mathbb
R}$ the following canonical transformation of the extended phase space
${\cal M}_E$
\ben
d\widetilde{t}&=&e^{\b(q)}\cdot dt\,,\nn\\
 \label{dtodah}\\
\widetilde{H}_\b\,&=&e^{-\b(q)}\cdot
 \Bigl(\,H+b_\b\,\Bigr)\nn
\en
maps the Toda lattice into the other integrable system. This canonical
transformation induces the following transformation of the Lax matrices
\bq
\widetilde{\cal L}_\b={\cal L}-\widetilde{H}_\b\cdot {e_\b}\,,
\qquad
\widetilde{\cal A}={e^{-\b(q)}}\cdot {\cal A} \,.\label{dtodal}
\eq
Here $H$, ${\cal L}$ and ${\cal A}$ are the Hamiltonian (\ref{todah})
and the Lax matrices (\ref{todal}) for the corresponding Toda lattice.
\end{prop}
In this proposition we explicitly determine the new parameter
$\widetilde{t}$ and the new associated  Hamilton function
$\widetilde{H}$. The corresponding spectral curves depend on the
choice of a representation of $\mathfrak g$. We prove that the Toda
lattice and the new integrable system relate to a common geometric
object in the one example only.

The number of the functional independent Hamilton functions
$\widetilde{H}_\b$, $\beta\in P$ depends upon the symmetries of the
associated root system. For closed Toda lattices associated with the
affine Lie algebras canonical time transformation has a similar form.
Similar canonical transformations may be applied to the relativistic
and discrete time Toda lattices.

We describe explicitly some new integrable systems related to the
standard three-particle Toda lattice and two-particle Toda lattices
associated to the affine algebras $X^{(1)}_2$. After an appropriate
point transformation of coordinates ( similar to (\ref{keplvar}), see
\cite{ts99c}) all the Hamilton functions have the common form
\bq
\widetilde{H}=p_x\,p_y+\dfrac{b}{xy}+
a\,x^{z_1}\,y^{z_2}+c\,x^{s_1}\,y^{s_2}+d\,,\qquad a,b,c,d\in{\mathbb
R}\,,
\label{tnewham}
\eq
where $z_{1,2}$ and $s_{1,2}$ are the roots of the different
quadratic equations and are related to the angles of the
corresponding Dynkin diagrams. Below we show these equations
explicitly
\ben
&A^{(1)}_3:\qquad &
\begin{array}{cc}
  z^2+3z+3=0 & s^2+3s+3=0 \\
\end{array}\nn\\
\nn\\
&B^{(1)}_2~C^{(1)}_2:\qquad &
\begin{array}{cc}
  z^2+4z+5=0 & s^2+4s+5=0 \\
  z^2+2z+2=0 & s^2+3s+5/2=0
\end{array}\nn\\
\nn\\
&D^{(1)}_2:\qquad &
\begin{array}{cc}
  z^2+2z+2=0 & s^2+2s+2=0 \\
  z^2+2z+2=0 & (s+2)^2=0
\end{array}\nn\\
\nn\\
&G^{(1)}_2:\qquad &
\begin{array}{cc}
  z^2+2z+4=0 & s^2+5s+7=0 \\
  z^2+2z+4=0 & s^2+3s+3=0 \\
  z^2+3z+7/3=0 & s^2+3s+3=0
\end{array}\nn
\en
Originally the integrable system with the Hamilton function
$\widetilde{H}$ (\ref{tnewham}) associated to the root system
$A^{(1)}_3$ was found by Drach \cite{dr35}.

The corresponding second integrals of motion $K$ are polynomials of
the third, fourth and sixth order in momenta. Note that for the
algebra $A^{(1)}_3$ all the three Hamiltonians $H_\b,~\b\in P$ are
equivalent. Two different Hamilton functions are associated with the
algebras $B^{(1)}_2$, $C^{(1)}_2$ and $D^{(1)}_2$. For the
$G^{(1)}_2$ algebra we have three independent potentials in
(\ref{tnewham}).

\section{On the common properties of the Toda lattices and the dual systems.}
\setcounter{equation}{0}
For the periodic Toda lattice associated with the root system $A_n$
the Hamilton function is
\bq
H(p,q)=
\dfrac12\,\sum_{i=1}^n \,p_i^2+a_i\,e^{q_i-q_{i+1}}\,,\qquad
a_i\in{\mathbb R}.\label{anh}
\eq
Here $\{p_i,q_i\}$ are canonical variables and the periodicity
conventions $q_{i+n}=q_i$ and $p_{i+n}=p_i$ are always assumed
for the indices of $q_i$ and $p_i$.

The exact solution of the equations of motion is due to existence of
a Lax equation with the following $n\times n$ Lax matrices
\cite{km75,fml76a}
\ben
{\cal L}^{(n)}(\mu)=\sum_{i=1}^n
p_i\,E_{i,i}&+&\sum_{i=1}^{n-1}\left(\,
e^{q_i-q_{i+1}}\,E_{i+1,i}+a_i\,E_{i,i+1}\right)+\nn\\ &+&
\mu\,e^{q_n-q_1}\,E_{1,n}+a_n\,\mu^{-1}\,E_{n,1}\,,\label{alax}
\en
\[
{\cal A}^{(n)}(\mu,q)=\sum_{i=1}^{n-1}\, e^{q_i-q_{i+1}}\,E_{i+1,i}+
\mu\,e^{q_n-q_1}\,E_{1,n}\,,
\]
where $E_{i,k}$ stands for the $n\times n$ matrix with unity on the
intersection of the $i$th row and the $k$th column as the only nonzero
entry.

According to \cite{ts99c,ts99d} canonical transformation of the
extended phase space (\ref{ttr}) by
\bq
v(p,q)=\exp(q_{j}-q_{j+1})\,,\qquad
\widetilde{H}=e^{q_{j+1}-q_{j}}\,(H-b)\,,\quad b\in{\mathbb R}
\label{trtan}
\eq
maps the Toda lattice into the dual integrable system with the
following equations of motion
\bq
\dfrac{dq_i}{d\widetilde{t}}=v^{-1}(q)\,\dfrac{dq_i}{dt}\,,\qquad
\label{dteq}
\dfrac{dp_i}{d\widetilde{t}}=
v^{-1}(q)\,\dfrac{dp_i}{dt}+\widetilde{H}\,(\delta_{i,j}-\delta_{i,j+1}\,)
\,.
\eq
Associated with the different indexes $j$ canonical mappings
(\ref{trtan}) are related with each other by canonical transformations
of the other variables $(p,q)$.

The mapping (\ref{trtan}) gives rise to the following transformation
of the Lax matrices
\bq
\widetilde{\cal L}(\mu)={\cal L}(\mu)-{\widetilde{H}}\,
E_{j,j+1}\,,\qquad\widetilde{\cal A}(\mu)=v^{-1}(q)\,{\cal A}(\mu)\,.
\label{antodal}
\eq
The corresponding transformation of the spectral curves is
\ben
{\cal C}:& -{\mu}-\dfrac{\prod^n_{i=1} a_i}{\mu} &=P(\l)
=\l^n+\l^{n-1}\,{\bf p}+
\l^{n-2}\,\left(\dfrac{{\bf p}^2}2-H\right)+\sum_{i=1}^{n-3} J_i\l^i
\nn\\
\nn\\
\label{todac}\\
\nn\\
\widetilde{\cal{C}}:&
 -\mu-
\dfrac{(a_j-\widetilde{H}) \prod_{i\neq j}^n a_i\,}{\mu} &
=\widetilde{P}(\l)=\l^n+\l^{n-1}\,{\bf p}+
\l^{n-2}\,\left(\dfrac{{\bf p}^2}2-b\,\right)
+\sum_{i=1}^{n-3} \widetilde{J}_i\l^i \,.\nn
\en
Here ${\bf p}=J_1=\sum p_i$ is the total momentum, $H$ and
$\widetilde{H}$ are the corresponding Hamilton functions and
$J_i,~\widetilde{J}_i$ are integrals of motion. Substituting the
fixed values of integrals of motion in the product of curves one can
construct the corresponding Lagrangian submanifold.

Applying Arnold's method \cite{arn89,fml76a} to the standard form of
the hyperelliptic curves $C$ and $\widetilde{C}$ (\ref{todac}) one
construct the action variables
\ben
{\mathfrak s}_i&=&\oint_{A_i} \dfrac12\left(\,
P(\l)+\sqrt{P(\l)^2-4\prod^n_{i=1} a_i\,}\right)\,d\l\,,\nn\\
\label{actoda}\\
\widetilde{\mathfrak s}_i&=&\oint_{\widetilde{A}_i}
\dfrac12\left(\,
\widetilde{P}(\l)+\sqrt{\widetilde{P}(\l)^2-4(a_j-\widetilde{H})
\prod_{i\neq j}^n a_i\,}\right)
\,d\l\,,\nn
\en
where $A_i$ and $\widetilde{A}_i$ are $A$-cycles of the Jacobi
variety of the algebraic curves (\ref{todac}), respectively
\cite{fml76a}. In fact the polynomials $P(\l),~\widetilde{P}(\l)$ and
$A$-cycles depend on the values of constants of motion, which are
dropped in the notation. The Abel transformation linearizes
equations of motion in terms of first kind abelian differentials on
the corresponding spectral curves.

Let parameters $a_i$ determine potential of the Toda lattice
(\ref{anh}) and parameters $\widetilde{a}_i$ and $b$ define the
potential of the dual system (\ref{trtan}). At the special choice of
the values of integrals of motion
\bq
H=b\,,\qquad \widetilde{H}=\widetilde{a}_j- \frac{\prod^n a_i}
{\prod_{i\neq j}^n \widetilde{a}_i}\,,
\qquad J_i=\widetilde{J}_i=\alpha_i
\label{tint}
 \eq
the initial curve ${\cal C}$ is equal to the resulting curve
$\widetilde{\cal C}$ (\ref{todac}). Thus, as for the Maupertuis
-Jacobi mapping \cite{bkf94}, integral trajectories of the Toda
lattice coincide with the trajectories of the dual system on the
intersection of the corresponding common levels of integrals ${\cal
M}_\alpha$ and $\widetilde{\cal M}_\alpha$. In the neighbourhood of
this intersection we can introduce the common set of the action
variables (\ref{actoda}) for the both systems. In this small
subvariety of the phase space the function $v(p,q)=v(\mathfrak s)$
depends on the action variables only.

At $a_i=1$ the Poisson bracket relations for the $n\times n$ Lax
matrices can be expressed in the $r$-matrix form
\[
\{\,{\on{\cal L}{1}}(\mu)\,,\,{\on{\cal L}{2}}(\nu)\,\}=
[\,r_{12}(\mu,\nu)\,,\,{\on{\cal L}{1}}(\mu)]+
[r_{21}(\mu,\nu)\,,\,{\on{\cal L}{2}}(\nu)\,]\,.
\]
Here we used the standard notations
\[{\on{\cal L}{1}}(\mu)= {\cal L}(\mu)\otimes I\,,
\qquad {\on{\cal L}{2}}(\nu)=I\otimes
{\cal L}(\nu)\,,\qquad r_{21}(\mu,\nu)=-\Pi\, r_{12}(\nu,\mu)\,\Pi\,,\]
and $\Pi$ is the permutation operator in ${\mathbb C}^n\times{\mathbb
C}^n$ \cite{ft87}. Canonical transformation of the extended phase space
(\ref{trtan}) maps the constant $r$-matrix for the Toda lattice
\[r_{12}(\mu,\nu)=r_{12}^{const}(\mu,\nu)=
\dfrac{1}{\mu-\nu}\,\left(\,\nu\sum_{m\geq
i}+\mu\sum_{m<i}\,\right)\, E_{im}\otimes E_{mi}\] into the following
dynamical $r$-matrix
\[\widetilde{r}_{12}(\mu,\nu)
=r_{12}^{const}(\mu,\nu)+\widetilde{\cal A}(\nu,q)\otimes E_{j,j+1}\,,
\]
where the second Lax matrix  $\widetilde{\cal A}(\nu,q)$ and,
therefore, the dynamical $r$-matrix $\widetilde{r}_{ij}(\mu,\nu)$
depend on coordinates only.

Another known $2\times 2$ Lax representation \cite{ft87,skl85a} for the
same Toda lattice is equal to
\bq
T(\l)=L_1(\l)\cdots L_n(\l)\,, \qquad
\dfrac{d\,T}{dt}=\Bigl[\,T(\l)\,,A_n(\l)\,\Bigr]\,,
\label{22toda}
\eq
where
\bq
L_i=\left(\begin{array}{cc}
 \l+p_i &\, e^{q_i} \\
 -a_{i-1}\,e^{-q_i}& 0
\end{array}\right)\,,\qquad
 A_i=\left(\begin{array}{cc}
 \l &e^{q_i} \\
 -a_i\,e^{-q_{i-1}}& 0
\end{array}\right)
\,.
\label{2alax}
\eq
Canonical transformation of the extended phase space (\ref{trtan})
gives rise to the following transformation of the Lax matrices
\ben
\widetilde{T}(\l)&=&
 L_1\cdots L_{j-1}\cdot\left[L_j\,L_{j+1}+
\left(\begin{array}{cc}H-b & 0 \\ 0 & 0\end{array}\right)\,\right]
\cdot L_{j+2}\cdots L_n
\,,\nn\\
\nn\\
\widetilde{A}_n(\l)&=&v^{-1}(q)\,A_n(\l)\,.\label{tr2l}
\en
At $a_i=1$ the Poisson brackets relations for the $2\times 2$ Lax
matrices $T(\l)$ (\ref{22toda}) satisfy the following Sklyanin
$r$-matrix relation
\bq
\{\,{\on{T}{1}}(\l)\,,\,{\on{T}{2}}(\nu)\,\}=
[\,R(\l-\nu)\,,\,{\on{T}{1}}(u)\,{\on{T}{2}}(\nu)\,]\,,\qquad
R(\l-\nu)=\dfrac{\Pi}{\l-\nu}\,.
\label{sklbr}
\eq
The mapping (\ref{trtan}) transforms these quadratic relations into
the following polylinear ones
\ben
\{\,{\on{\widetilde{T}}{1}}(\l)\,,\,{\on{\widetilde{T}}{2}}(\nu)\,\}&=&
[\,R(\l-\nu)\,,\,{\on{\widetilde{T}}{1}}(\l)\,
{\on{\widetilde{T}}{2}}(\nu)\,]\nn\\ &+&
[\,r_{12}^{dyn}(\l,\nu)\,,\,{\on{\widetilde{T}}{1}}(\l)\,]+
[\,r_{21}^{dyn}(\l,\nu)\,,\,{\on{\widetilde{T}}{2}}(\nu)\,]
\,.\nn
\en
The corresponding dynamical $r$-matrix is given by
\[
r_{12}^{dyn}(\l,\nu)=A_n(\l,q)\otimes\,\left(\, L_1\cdots L_{j-1}\cdot
\left(\begin{array}{cc}
 1 & 0 \\
 0 & 0
\end{array}\right)
\cdot L_{j+1}\otimes L_n\,\right)\,.
\]
Here all matrices $L_k$ depend on the spectral parameter $\nu$ and
$A_n(\l,q)$ is the second Lax matrix (\ref{22toda}).

Now we look at the separation of variables in the framework of the
traditional consideration of the Toda lattice. A complete list of
references can be found in \cite{km75,fml76a,skl85a}. Below we put
$a_i$=1 and $j=1$ without loss of generality, so that
\[
\widetilde{H}=\exp(q_2-q_1)\,(H+b)\,,\qquad
\widetilde{T}=
 \left[L_1\,L_{2}+
\left(\begin{array}{cc}H+b & 0 \\ 0 & 0\end{array}\right)\,\right]
\cdot L_{3}\cdots L_n\,.
\]
This transformation changes the first row of the Lax matrix $T(\l)$
only
\bq
\label{2ttr}\\
T=\left(\begin{array}{cc} {\mathbb A}(\l) & {\mathbb B}(\l)
\\ {\mathbb C}(\l) & {\mathbb D}(\l)
\end{array}\right)\quad \mapsto\quad
\widetilde{T}=\left(\begin{array}{cc}
\widetilde{\mathbb A}(\l) & \widetilde{\mathbb B}(\l) \\
{\mathbb C}(\l) & {\mathbb D}(\l)
\end{array}\right)
\,.
\eq
The separated variables $\{\l_1\,\l_2\,\ldots,\l_{n-1}\}$ for both
systems are zeroes of the nondiagonal common entry
\bq
{\mathbb C}(\l)=
\gamma\cdot \prod_{i=1}^{n-1}(\l-\l_i)\,.
\label{sepv}
\eq
An additional set of variables is defined by the second common entry
\[\mu_i={\mathbb D}\,(\l_i)\, \qquad\mbox{\rm such that}\qquad
\{\l_i,\log\mu_k\}=\delta_{ik}\quad i,k=1,\ldots, n-1\,.
\]
From $\det T(\l)=1$ and $\det \widetilde{T}(\l)=(1-\widetilde{H})$
one immediately obtains the one-dimensional equations
\bq
\begin{array}{lr}
  {\mathbb A}(\l_i)=\mu_i^{-1} &\qquad \mu_i+\mu_i^{-1}=P(\l_i)\,, \\
  \\
  \widetilde{\mathbb
A}(\l_i)=(1-\widetilde{H})\mu_i^{-1} &\qquad
\mu_i+(1-\widetilde{H})\mu_i^{-1}=\widetilde{P}(\l_i)\,.
\end{array}
\label{sepeqt}
\eq
At the special choice of parameters and values of integrals
(\ref{tint}) initial separated equations coincide with the resulting
ones.

Thus we prove that the initial and resulting systems have a common
set of separated variables. On the other hand canonical
transformation of the extended phase space (\ref{trtan}) changes the
form of the B\"acklund transformation \cite{ts99d} and the
bihamiltonian structure, which are known for the Toda lattice.

Having obtained a simple change of the separated equations
(\ref{sepeqt}), one can hope that there is also a simple modification
of the one-dimensional Baxter equations in quantum mechanics. Recall
that from the works of Sklyanin \cite{skl85a,khl99} one knows that
the eigenfunctions of the quantum Toda lattice Hamiltonian are given
by
\[\psi_E(q)=\int\,C(\lambda,E)\,\psi_\lambda(q)\, d\lambda\,,
\qquad C(\lambda,E)=\prod^{n-1}_{j=1} c(\lambda_j,E)\,.\]
Here $\psi_\lambda$ are renormalized Whittaker functions and the
functions $c(\lambda,E)$ satisfy to one-dimensional Baxter equation
\[P(\lambda)\,c\,(\lambda,E)=i^n\,c\,(\lambda+i\hbar,E)
+i^{-n}\,c\,(\lambda-i\hbar,E)\,,
\]
where $P(\lambda)$ is a trace of the quantum monodromy matrix
$T(\l)$. In the classical limit the polynomial $P(\lambda)$ enters in
the spectral curve (\ref{todac}).

By using a similar approach \cite{skl85a,khl99} we can suppose that
the eigenfunctions for the dual system are expressed in terms of the
same Whittaker functions
\[\widetilde{\psi}_{\widetilde E}\,(q)=\int\,\widetilde{C}(\lambda,\widetilde{E})\,
\psi_\lambda(q)\, d\lambda\,,\]
whereas the corresponding one-dimensional Baxter equation has to be
changed
\[\widetilde{P}(\lambda)\,\widetilde{c}\,(\lambda,\widetilde{E})=
i^n\,\left(1-\widetilde{E}\right)\,\,\widetilde{c}\,(\lambda+i\hbar,\widetilde{E})
+i^{-n}\,\widetilde{c}\,(\lambda-i\hbar,\widetilde{E})\,,
\]
in accordance with the corresponding classical separated equations
(\ref{sepeqt}). In the classical limit the polynomial
$\widetilde{P}(\lambda)$ enters in the spectral curve (\ref{todac}).

Recall that in the Birman-Schwinger formalism we can estimate
spectrum and eigenvalues of the one Hamiltonian $\widetilde{H}$ by
using known spectrum and eigenvalues of the dual Hamiltonian $H$. So
it is interesting to study such a duality in framework of the quantum
${\mathbb Q}$-operator theory as an example for the Toda lattice.

\section{The Goryachev-Chaplygin top}
\setcounter{equation}{0}
The construction of canonical transformations of the extended phase
space proposed for the St\"ackel systems was inspired by the Kepler
and Liouville results. These transformations give rise to the shift
of the Lax matrices (\ref{ltr}) and it allowed us to introduce an
integrable system dual to the Toda lattice. As above we can try to
construct another integrable systems starting with known ones by
using the mapping of the $2\times 2$ Lax matrices for the Toda
lattice (\ref{tr2l}).

We start with any Lax matrix $T(\l)$ in the form (\ref{22toda}).
Substituting the known Hamilton function $H$ into the mapping
(\ref{tr2l}) one calculates a new matrix $\widetilde{T}$ and a  new
Hamilton function, which may be tested on integrability. Note that to
construct new Lax matrices by the rule (\ref{ltr}-\ref{dtodal}) we
have to predict the new Hamiltonian $\widetilde{H}$ from some
external reasons.

As an example here we consider the Goryachev-Chaplygin top. We
introduce coordinates on the dual space to the Lie algebra $e(3)$
with the standard Lie-Poisson brackets
\ben
&&\bigl\{ l_i\,, l_j\,\bigr\}=\varepsilon_{ijk}\,l_k\,,
\qquad
\bigl\{ l_i\,, g_j\,\bigr\}=
\varepsilon_{ijk}\,g_k\,,
\nn\\ \label{e3}\\
&&\bigl\{ g_i\,, g_j\,\bigr\}= 0\,, \qquad i, j, k=1, 2, 3.
\nn
\en
The orbits on $e(3)^*$ are fixed by values of the two  Casimir
operators $C_1=(g, g);~C_2=(l, g)$. The Hamilton function for the
Goryachev-Chaplygin top is equal to
\bq
H=\dfrac12\left(l_1^2+l_2^2+4l_3^2\right)-p\,l_3+g_2\,,\qquad p\in
{\mathbb R}
\label{hgc}
\eq
It is a completely integrable top on the one-parameter subset of
orbits ${\cal O}$ ($C_1=const\,,~ C_2=0$) in $e(3)^*$. The
corresponding $2\times 2$ Lax matrix $T(\l)$ was obtained by
Sklyanin \cite{skl85}. According to \cite{ts95} this matrix is
closely related to the Lax matrix for the three-particle Toda
lattice and it may be factored as
\[T(\l)=T_1(\l)\, T_{23}(\l)\,,\quad\mbox{\rm where}\quad
T_1=\left(\begin{array}{cc} \l-p+2l_3& e^{q}\\
-e^{-q}&0\end{array}\right)\,,\] and
\bq
T_{23}=\left(\begin{array}{cc} \l^2-2l_3\l-l_1^2-l_2^2
\,&ie^{q}[\l(g_1-ig_2)-g_3(l_1-il_2)]
\\ ie^{q}[\l(g1+ig_2)-g_3(l_1+il_2)]
\,&g_3^2\end{array}\right)\,.
\label{gclax}
\eq
By using the transformation of the Lax matrices (\ref{tr2l}) proposed
for the Toda lattices one constructs another Lax matrices
\[\widetilde{T}(\l)=T_1\cdot\left[\,T_{23}+
\left(\begin{array}{cc} H-b&0\\0&0\end{array}\right)\,\right]\,,
\qquad \widetilde{A}=v^{-1}\,A
\]
which describes a new integrable system on the same one-parameter
subset of orbits $\cal O$. So the canonical transformation of the
extended phase space (\ref{ttr}) by $v=g_3^{-2}$, i.e.
\bq
\widetilde{H}=g_3^{2}\,(H-b)\,, \label{tgch}
\eq
preserves integrability. Moreover, initial and resulting systems are
separable in the common system of separated variables. By using
Maupertuis' principle a similar transformation of the extended phase
space (\ref{tgch}) has been obtained in \cite{sel98}.

Transformation of the corresponding spectral curves has the expected
form
\ben
&{\cal C}:\qquad&
\mu-\dfrac{\l^2(g,g)}{\m}=\l^3-p\,\l^2-2H\,\l-K\,,
\nn\\
&\widetilde{\cal C}:\qquad&
\mu-\dfrac{\l^2(g,g)+2\widetilde{H}}{\m}=\l^3-p\,\l^2-2b\,\l-\widetilde{K}\,.\nn
\en
As above, these integrable systems are topologically equivalent.

Starting from the matrix $T_{23}(\l)$ (\ref{gclax}) we can construct
$2\times 2$ Lax matrix for the so-called
Ko\-wa\-lev\-ski-Goryachev-Chaplygin top (see references within
\cite{ts95}). This Lax matrix is closely related to the Lax matrix
for the Toda lattice associated with the root system $BC_2$.
Starting with the induced transformation of the $2\times 2$ Lax
matrices for the $BC_n$ Toda lattices we can construct a  new
integrable system related to the Kowalevski-Goryachev-Chaplygin top.
Similar transformations of the extended phase space have been
proposed in \cite{sel00} directly from  Maupertuis' principle.

\section{Conclusion}
The modest aim of this review was to collect some old and new
examples of  canonical transformations of the extended phase
space, which map a given integrable system into the other
integrable system.

The Sections 4-10 together show that all the examples have many
common properties. Analysis of these common properties could allow
us to join different integrable systems into the classes of
topologically equivalent systems and to study these classes instead
of considering of the individual systems. Each of this class of
integrable system may be related to the class of topologically
equivalent $n$-dimensional Lagrangian submanifolds, which are
diffeomorphic to the $n$-dimensional torus. In this approach the
different integrable systems are associated with the various
parametric forms of the common integrable manifold.

It remains unclear how to construct canonical transformations of the
extended phase space ${\cal M}_E$ for a given integrable system.
Moreover, up to now one does not know all consequences of the action of
canonical transformations on the St\"ackel systems, Toda lattices and
other known integrable systems.

There is still much to do: to describe modification of
bi-hamiltonian structures and the B\"acklund transformations for
the St\"ackel systems and the Toda lattices, to transform the
corresponding stationary flows of hierarchies of nonlinear
evolution equations and to consider composed transformations
similar to the Kolosoff mapping for the Kowalevski top.

For all the examples, after canonical transformations of ${\cal
M}_E$, one usually gets dynamical $r$-matrices instead of  constant
ones. We have to check that these resulting $r$-matrices satisfy to
the classical dynamical Yang-Baxter equation and to interpret
properly these dynamical matrices in a general theory of dynamical
$r$-matrices.

\section{Acknowledgement}
I thank I.V. Komarov for comments and discussions. This work was
partially supported by RFBR grant 99-01-00698 and by INTAS grant
99-01459.


\end{document}